\documentclass[14pt,a4paper]{extarticle}
\usepackage[english]{babel}
\usepackage{graphicx}
\usepackage{geometry}
\usepackage{amsmath}
\usepackage{multirow}
\usepackage{multicol}
\usepackage{caption}
\usepackage{float}
\usepackage{setspace}
\usepackage{csquotes}
\usepackage[
backend=biber,
style=gost-numeric,
sorting=none,
language=auto,
]{biblatex}

\begin{document}
\title{Compton suppressed scintillation spectrometer based on a cluster of 9$\times$CeBr$_{3}$-NaI(Tl) phoswich detectors}
\date{}
\maketitle
\begin{center}
	\author{M.A. Povolotskiy$^{a,}$\footnote{E-mail: povolotskiy@jinr.ru},
		Yu.G. Sobolev$^{a,}$\footnote{E-mail: sobolev@jinr.ru},
		S.S. Stukalov$^{a,b,}$\footnote{E-mail: stukalov@jinr.ru}}
\end{center}
\begin{center}
	\
	\\{$^{a}$\,Joint Institute for Nuclear Research, Dubna, Russia}
	\\{$^{b}$\,Voronezh State University, Voronezh, Russia}
\end{center}

\begin{abstract}
	
	Measurements of the characteristics of an Compton suppressed spectrometer consisting of 9$\times$CeBr$_{3}$-NaI(Tl) phoswich detectors surrounded by four CsI(Tl) scintillation detectors intended for suppressing the Compton component of the spectrum have been carried out. Measurements were performed using gamma and neutron radiation sources. Key parameters of the spectrometer have been determined: the suppression factor of the Compton part of the $\gamma$-spectrum, as well as the dependencies of the neutron detection efficiency on their energy at various detection threshold values.
	
	\vspace{0.2cm}
\end{abstract}
\vspace*{6pt}

\noindent
PACS: 07.85.Nc, 29.40.Mc, 29.30.Hs

\label{sec:intro}

\section*{\textbf{Introduction}}

Clusters of phoswich detectors, consisting of CeBr$_3$-NaI(Tl) or LaBr$_3$-NaI(Tl) scintillators, are actively used in experiments studying the direct $\gamma$-decay processes of Giant or Pygmy Dipole Resonances (GDR, PDR) formed in reactions with charged particles \cite{wasilewska2022gamma}, as well as in studies of $\beta$-decay processes of radioactive nuclei \cite{li2025beta}.

The application of phoswich detectors in modern nuclear physics experiments is due to their ability to efficiently separate high-energy $\gamma$-quanta from cascades of low-energy $\gamma$-radiation of high multiplicity.

This is especially important when analyzing events of nuclear reactions in which the total energy of the registered $\gamma$-radiation cascades is close in value to the energies of high-energy $\gamma$-quanta.

Phoswich detectors based on modern scintillators such as CeBr$_3$, LaBr$_3$, GaGG:Ce, etc. \cite{moses2005potential, yeom2013first, quarati2013scintillation}, have a high energy and time resolution. 
This makes them promising for applications in precision $\gamma$-spectroscopy tasks.

The high values of light output and short scintillation pulse-time of these materials, combined with the average values of the mass numbers of the nuclei composing them, also allow considering the possibility of their use as detectors for fast neutrons $E\sim 1 \div 50$ MeV.

This work is dedicated to the issues of applying a cluster consisting of 9$\times$CeBr$_3$-NaI(Tl) phoswich detectors as an Compton suppressed $\gamma$-spectrometer, as well as a detector for fast neutrons.

The spectrometer is one of the elements of the MULTI setup \cite{sivavcek2020multi}, created at the Flerov Laboratory of Nuclear Reactions (FLNR), JINR, Dubna, and will be used both in conjunction with the 4$\Pi$-spectrometer MULTI, see Fig. \ref{fig:1}, and independently.

\begin{figure} [H]
\centering
\includegraphics[width=0.5\linewidth]{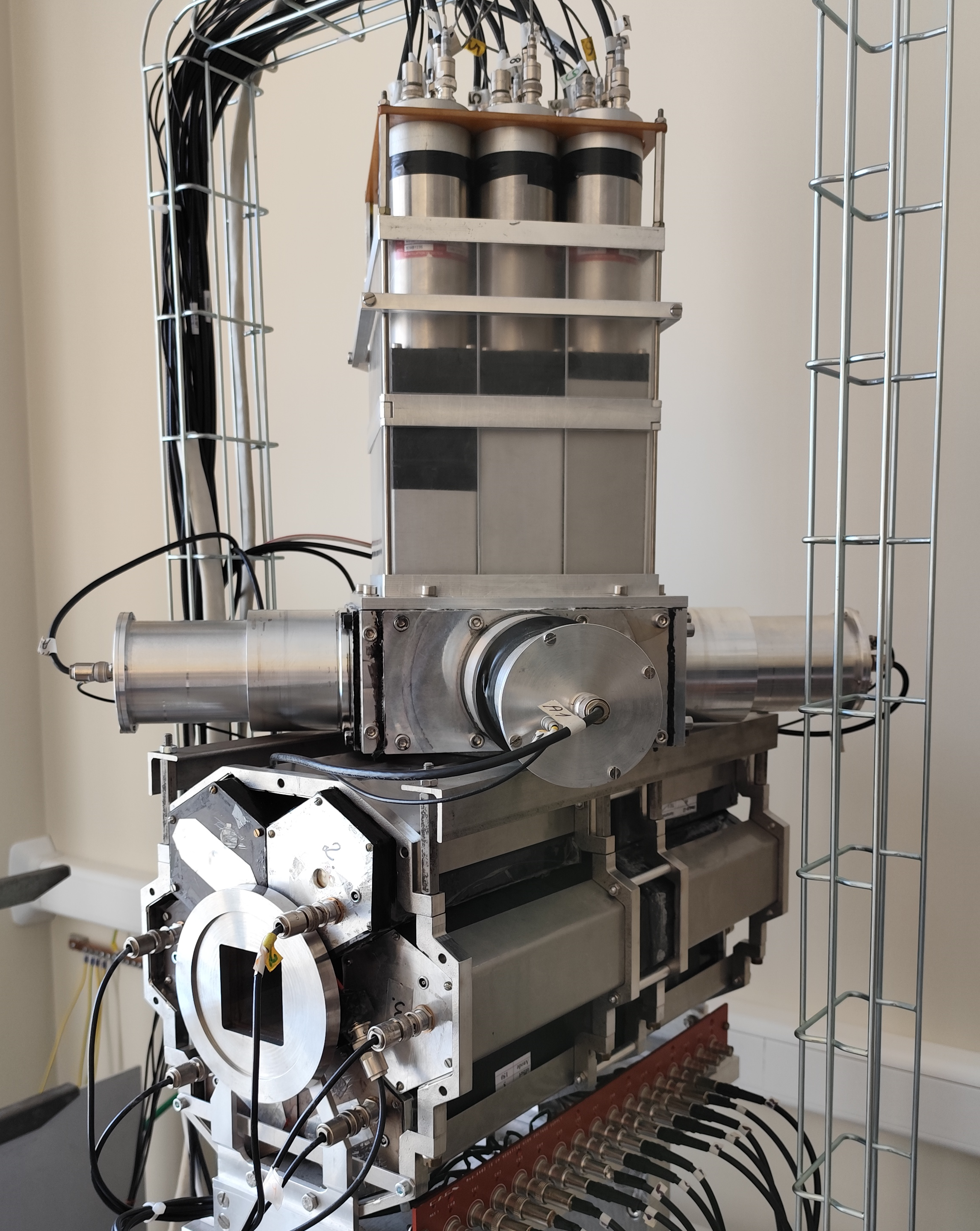}
\caption{\label{fig:1}\it{Photo of the MULTI experimental setup.}}
\label{fig:1}
\end{figure}

The spectrometer is intended for measuring total reaction cross sections with neutron-rich nuclei located far from stability line, as well as for studying their $\beta$-decay processes.
\section*{\textbf{The cluster of CeBr$_3$-NaI(Tl) phoswich detectors as a compton suppressed spectrometer}}

\begin{figure} [H]
	\centering
	\includegraphics[width=0.8\linewidth]{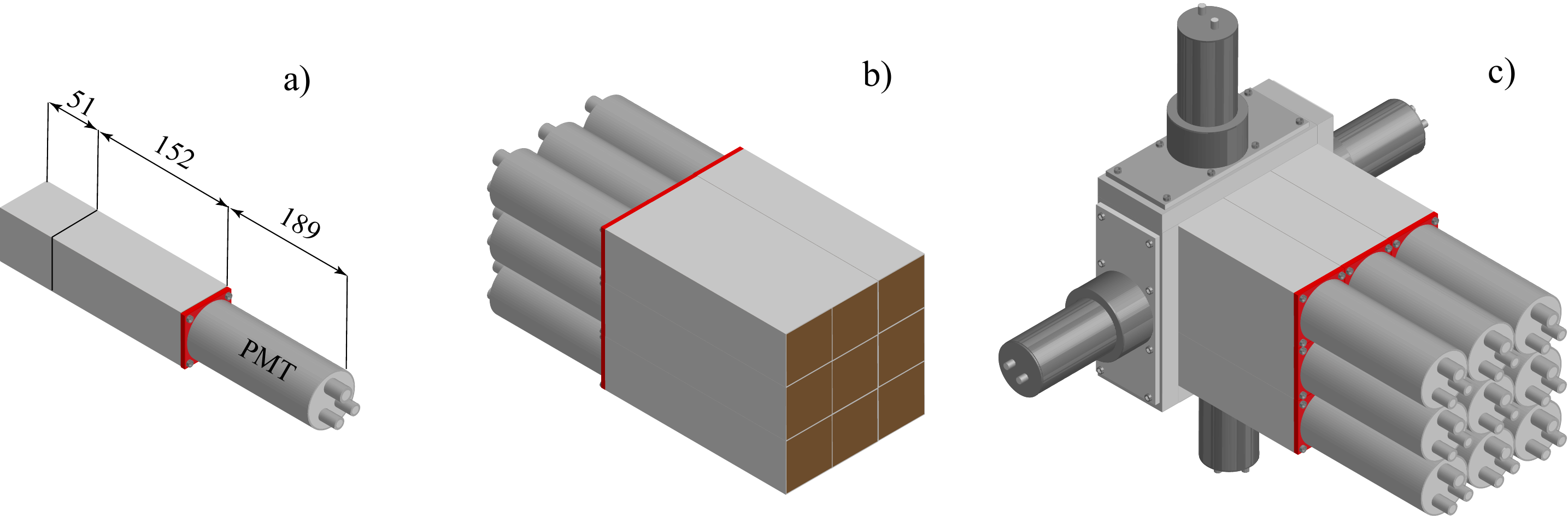}
	\caption{\it{a) Scintillation CeBr$_3$-NaI(Tl) phoswich detector. b) Cluster of 9$\times$phoswich detectors. c) Compton suppressed $\gamma$-spectrometer from a cluster of phoswich detectors and 4 suppressed Compton shield detectors.}}
	\label{fig:2}
\end{figure}

The cluster is a close-packed assembly of nine phoswich detectors arranged in a 3$\times$3 matrix pattern, see Fig.\ref{fig:2}b).

A phoswich detector (see Fig.\ref{fig:2}a)) consists of a pair of CeBr$_3$ and NaI(Tl) scintillation crystals with dimensions of 50$\times$50$\times$50 mm$^3$ and 50$\times$50$\times$150 mm$^3$, respectively, optically connected and placed in a thin-walled sealed housing with a transparent end connected to a photomultiplier. 

The thickness of the CeBr$_3$ crystal was chosen to absorb most part  of low-energy $\gamma$-quanta ($E\gamma$ less than 1 MeV) in the scintillator material, while a high-energy $\gamma$-quanta ($E\gamma$ more than 5 MeV) passed through the CeBr$_3$ crystal and had a first interaction with atoms of the NaI(Tl). GEANT-4 calculations showed that the probability of $\gamma$-quanta with $E\gamma$ = 1 MeV passing through the CeBr$_3$ material into the NaI(Tl) scintillator does not exceed $\sim$10$\%$.

\begin{figure} [H]
\centering
\includegraphics[width=1\linewidth]{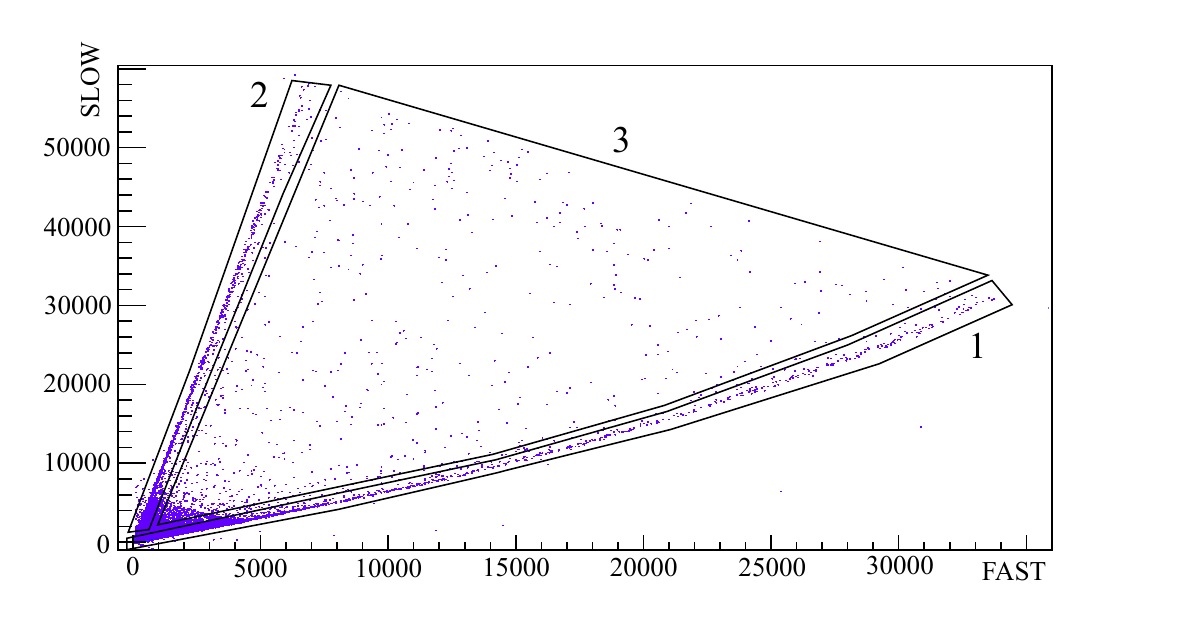}
\caption{\it{Two-dimensional spectrum of the distribution of PMT pulse charges, registering scintillations of CeBr$_3$-NaI(Tl) crystals of a phoswich detector from a $^{60}$Co source. "1" – area of events of $\gamma$-quantum registration only in CeBr$3$, "2" – area of events of $\gamma$-quantum registration only in NaI(Tl), "3" - area of events of $\gamma$-quantum registration in both CeBr${3}$ and NaI(Tl) scintillators}}
\label{fig:3}
\end{figure}

The CeBr$_3$ crystal has a single-component scintillation, characterized not only by a fast scintillation rise time ($\sim$10 ns) but also a short decay time ($\sim$50 ns) \cite{qin2020characteristics}, as well as high light output, comparable in magnitude to NaI(Tl). These parameters combine excellently with the parameters of the slow NaI(Tl) scintillator within the phoswich detector to ensure effective pulse shape analysis.
Fig. \ref{fig:3} shows a two-dimensional spectrum of the distribution of the photomultiplier anode signal charges from the fast and slow scintillation components during the registration of $\gamma$-quanta from a $^{60}$Co source by the CeBr$_3$-NaI(Tl) phoswich detector.

The X-axis "FAST" represents the charge of the photomultiplier (PMT) anode pulses measured in a short time interval $\tau_{FAST}$ = 30 ns. This time interval corresponds to ~60$\%$ of the CeBr$3$ scintillation. 

The Y-axis "SLOW" represents the magnitude of the pulse charge integrated in the time gate $\tau_{SLOW}$ = 400 ns, which follows the "FAST" interval. In this time gate, mainly the slow component of the NaI(Tl) scintillation was integrated. 

Points populating the lower diagonal area "1" correspond to events of $\gamma$-quantum registration only in the CeBr$_3$ scintillator material. Points populating the area marked by contour "2" correspond to acts of $\gamma$-quantum registration only in the NaI(Tl) scintillator material. The area marked by contour "3", located between contours "1" and "2", is populated by points related to events in which the $\gamma$-quantum and its secondary interaction products, formed during the passage through the materials, were registered in both scintillators.
Analysis of the PMT pulse shape of the phoswich detector using the two-dimensional distribution described above allows for the separation of low-energy and high-energy $\gamma$-quanta, as well as the implementation of the Compton suppressed spectrometer mode.

 We can make a conclusions about the complete absorption of the $\gamma$-quantum in the CeBr$_3$ crystal when two following conditions are simultaneously met, based on the analysis of $\gamma$-quantum registration events by the phoswich detector using the two-dimensional spectrum (Fig. \ref{fig:3}) and spectra in the surrounding detectors:

-absence of signals in neighboring phoswich detectors and CsI(Tl)  Compton shield detectors;

-absence of the slow scintillation component corresponding to energy deposition in the NaI(Tl) part of the phoswich detector.

\section*{\textbf{Measurement of the Compton Component Suppression Factor}}

The study of the characteristics of the Compton suppressed spectrometer based on the cluster of phoswich detectors  was carried out using $\gamma$-sources: $^{60}$Co, $^{137}$Cs, and $^{152}$Eu.

Measurements of Compton suppression coefficient AC (see below) (which is a ratio of summ of the events in Compton-part of energy spectra measured in inclusive and Compton suppression modes) were performed in $\gamma$-$\gamma$ coincidence using the $^{60}$Co source. The trigger detector was a CeBr$_3$ scintillator. Its logical signal initiated event recording in the VME data acquisition system \cite{ssyl4}.

The Mesytec MVLC crate controller \cite{ssyl4} received a start pulse from the trigger detector. Then the MDPP16-QDC \cite{ssyl4} system recorded signals from all detectors of the cluster within a given time gates.

To select $\gamma$-$\gamma$ coincidences, events were chosen under the conditions in which:

- the fast time pulses from the trigger and phoswich detectors are in coincidence in narrow ($\Delta T_{FWHM}$=10ns) time window. 

- the trigger CeBr$_3$ detector registered a total absorbed $\gamma$-quantum $E{\gamma}$=1332 keV, and the cluster registered a $\gamma$-quantum with energy $E_{\gamma}$=1173 keV. 

This method effectively suppresses background from the intrinsic glow of the CeBr$_3$ detector.

 Event selection algorithm was applied aimed at identifying the complete absorption of the $\gamma$-quantum in its CeBr$_3$ part of phoswich detector. 
 The key criterion was the requirement that the $\gamma$-quantum energy be registered only in the CeBr$_3$ part of the phoswich detector, with a complete absence of signals in its NaI(Tl) section, as well as in neighboring detectors of the cluster and the CsI(Tl) block. This allows for the selection of events in which the Compton-scattered $\gamma$-quantum does not leave the volume of the CeBr$_3$ crystal.

The value of measured Peak efficiency for $\gamma$-quanta $E\gamma$ = 1.17 MeV (in the full absorption peak) for the CeBr$_3$ part of the cluster of phoswich detector, located at 10 cm from the $^{60}$Co source, was $\varepsilon$= 21.8\%. 
value of Peak efficiency is more than 21 times higher than one for HPGe ($\O$ = 55 mm, $H$ = 60 mm) detectors \cite{challan2013gamma} located at the same distance.

Measurements of Compton suppression coefficient with the $^{152}$Eu $\gamma$-source were carried our in "self-trigger" mode.  

$^{152}$Eu $\gamma$-source emits the cascades of $\gamma$-quanta in a wide energy range $E_{\gamma}$= 121 $\div$ 1408] keV and can be used for a close approach to a real experiment on nuclear reactions, where it is impossible to unambiguously separate the Compton continuum.

\begin{figure}[H]
\centering
\begin{minipage}{0.45\textwidth}
	\centering
	\includegraphics*[width=\textwidth]{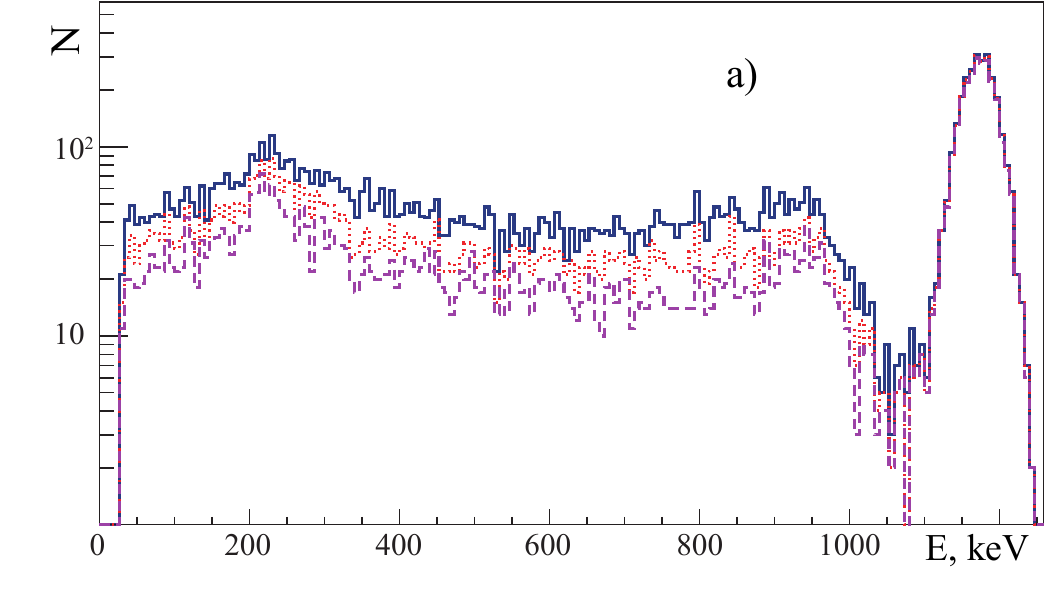}
\end{minipage}
\hfill
\begin{minipage}{0.45\textwidth}
	\centering
	\includegraphics*[width=\textwidth]{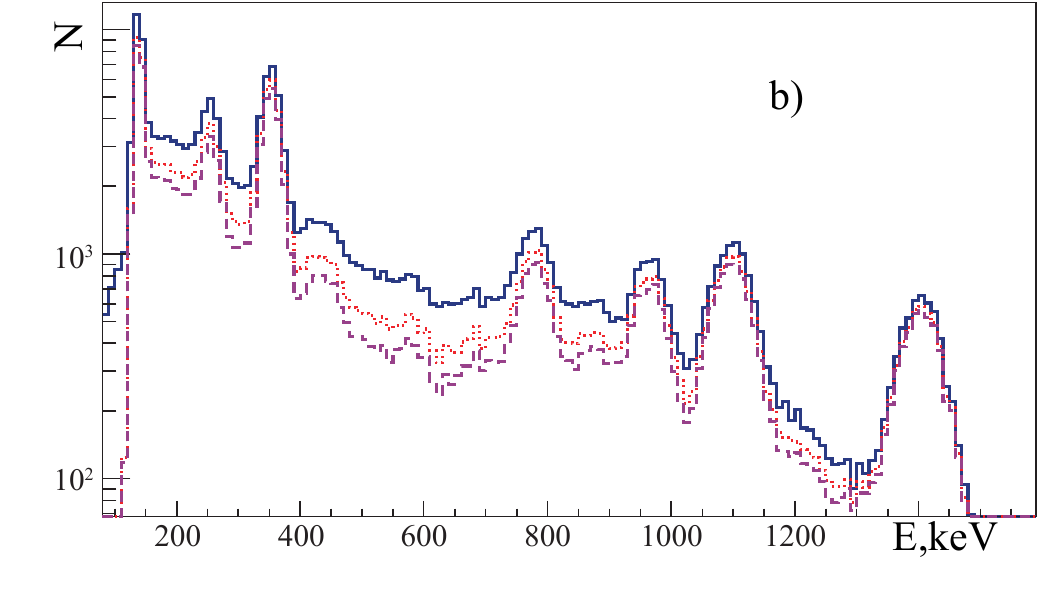}
\end{minipage}
\caption{\it{Energy spectra of $\gamma$-sources $^{60}$Co (a) and $^{152}$Eu (b) with superimposed amplitude spectra obtained under Compton suppressed shield conditions.}}
\label{fig:4}
\end{figure}

Fig.\ref{fig:4} shows examples of energy spectra of $\gamma$-sources $^{60}$Co (a) and $^{152}$Eu (b), measured by one of the phoswich detectors. Solid lines represent the inclusive energy spectrum. Dashed lines represent the spectrum obtained taking into account scattering into neighboring phoswich detectors. Dash-dotted lines represent the spectrum with additional consideration of scattering into the CsI(Tl) detector block. The energy threshold for Compton scattering registration was set at 60 keV.

Data analysis with $^{152}$Eu shows a loss of events in the full absorption peaks was observed — a known effect for Compton suppressed spectrometers. The loss of events arises due to random coincidences of two $\gamma$-quanta from one cascade, registered in neighboring detectors within the time window allotted for Compton scattering registration\cite{britton2014monte}.

The reason lies in a fundamental limitation of the method: the electronic system cannot distinguish between a pair of signals from two cascade $\gamma$-quanta registered in different detectors, and a pair from one $\gamma$-quantum that underwent Compton scattering. This leads to a systematic loss of full absorption events. The severity of the effect is directly influenced by two factors: the presence of cascade transitions in the source and its intensity.

For monoenergetic sources, such as $^{137}$Cs, or for correlated $\gamma-\gamma$ measurements with $^{60}$Co, the shape of the spectrum allows clear separation of the energy region corresponding to Compton scattering events.

In these cases, the suppression factor AC is calculated as the ratio of summ of the events in Compton-part of energy spectra measured in inclusive and Compton suppression modes:
\begin{equation}
	\label{eq:1}
	AC=(1-\frac{N_{AC}}{N})\cdot100\%
\end{equation}

where $N$ and $N_{AC}$ are the number of events in the Compton region before and after applying the suppression algorithm, respectively.
It is is difficult to isolate a pure Compton region of the spectrum for sources emitting cascades of $\gamma$-quanta in a wide energy range (e.g., $^{152}$Eu) due to the overlap of numerous photo-peaks and their Compton continua. The Compton Suppression Factor (CSF) criterion "peak-to-total" \cite{britton2014monte} was applied for analyzing spectra from $^{152}$Eu, which evaluates suppression effectiveness by the reduction of background under full absorption peaks. The calculation was performed using the following formula:
\begin{equation}
	CSF_{total}=\left(\frac{N_{peak}}{N_{total}}\right)_{w/o.AC}/\left(\frac{N_{peak}}{N_{total}}\right)_{AC}
\end{equation}
where $\left(\frac{N_{peak}}{N_{total}}\right)_{w/o.AC}$ is the ratio of the number of events in the peak without substrate to the total number of events (total) without applying suppression, $\left(\frac{N_{peak}}{N_{total}}\right)_{AC}$ is the ratio of the number of events in the peak without substrate to the total number of events (total) with suppression applied.

Table 1 presents the AC coefficients obtained from the analysis of spectra of neighboring phoswich detectors (excluding spectra from the 4$\times$CsI(Tl) detectors) as an example of the obtained measurement results.

Additionally, Table 1 presents the AC(CsI) suppression coefficients of the Compton part of the spectrum taking into account analysis in all detectors of the spectrometer to evaluate the effectiveness of suppressing the Compton part of the spectrum by the CsI(Tl) detectors. It is seen from Table 1 that including an additional Compton shield of 4$\times$CsI(Tl) detectors increases the effectiveness of suppressing the Compton component of the spectrum by $\times$1.5 times.
\begin{table}[H]
	\centering
	\label{tab:suppression_results}
	\renewcommand{\arraystretch}{1.2}
	\small
	\begin{tabular}{lcc}
		\hline
		Detector & AC & AC(CsI) \\
		&  (\%) & (\%) \\
		\hline
		Phoswich 1 &  31.0 $\pm$ 1.2& 47.0 $\pm$ 1.0\\
		Phoswich 2 &  44.4 $\pm$ 0.9& 53.3 $\pm$ 0.8\\
		Phoswich 3 &  32.3 $\pm$ 1.2& 45.4 $\pm$ 1.0\\
		Phoswich 4 &  46.2 $\pm$ 0.9& 58.0 $\pm$ 0.8\\
		Phoswich 5 &  63.8 $\pm$ 0.7& 65.6 $\pm$ 0.6\\
		Phoswich 6 &  46.3 $\pm$ 1.0& 55.1 $\pm$ 0.9\\
		Phoswich 7 &  31.3 $\pm$ 1.3& 50.4 $\pm$ 1.0\\
		Phoswich 8 &  44.7 $\pm$ 1.0& 55.8 $\pm$ 0.9\\
		Phoswich 9 &  32.8 $\pm$ 1.2& 49.1 $\pm$ 1.0\\
		\hline
	\end{tabular}
	\caption{\it{Suppression factors of the Compton part of the spectrum obtained in measurements with $^{60}$Co taking into account only the cluster of phoswich detectors (AC) and suppression by the cluster supplemented with a block of 4$\times$CsI(Tl) detectors (AC(CsI)).}}
\end{table}

Results of measurements with subsequent processing for the $\gamma$-source $^{152}$Eu are also presented. Due to the large volume of data because of the need to calculate suppression for the peak corresponding to each energy of $\gamma$-quanta emitted by the source, results for one detector of the cluster are presented as an example.

\begin{table}[H]
	\centering
	\label{tab:csf_values}
	\renewcommand{\arraystretch}{1.2}
	\small
	\begin{tabular}{lccc}
		\hline
		$E_{\gamma}$ & $(N_\text{peak}/N_\text{tot})_{w/o AC}$ & $(N_\text{peak}/N_\text{tot})_{AC}$ & CSF \\
		(keV) & $\times 10^{-3}$ & $\times 10^{-3}$ & \\
		\hline
		245 & 28.3$\pm$0.3 & 36.3$\pm$0.5 & 1.282$\pm$0.026 \\
		344 & 89.9$\pm$0.7 & 126.1$\pm$1 & 1.402$\pm$0.016 \\
		779 & 18.4$\pm$0.3 & 25.6$\pm$0.5 & 1.391$\pm$0.035 \\
		964 & 14.7$\pm$0.3 & 21.4$\pm$0.4 & 1.455$\pm$0.04 \\
		1112 & 25.8$\pm$0.3 & 35.9$\pm$0.5 & 1.391$\pm$0.029 \\
		1408 & 21.1$\pm$0.3 & 30.2$\pm$0.5 & 1.431$\pm$0.033 \\
		\hline
	\end{tabular}
	\caption{\it{Suppression coefficients of the Compton part of the spectrum obtained in measurements with the $^{152}$Eu source with intermediate calculated peak/total values. }}
\end{table}

The comparison with modern detector systems allows highlighting the main advantages of the developed spectrometer, the basis of which is a cluster of CeBr$_3$-NaI(Tl) phoswich detectors with an Compton suppressed shield of four CsI(Tl) detectors.

Although the combination of an HPGe–BGO Compton suppressed clover detector remains a common solution for precision $\gamma$-spectroscopy and provides high-quality Compton background suppression, its drawbacks — low full absorption peak registration efficiency relative to CeBr$_3$ \cite{challan2013gamma}, limited HPGe detector volume, and operational complexity — make it less suitable for studying reactions with light nuclei. In such tasks, where high-energy $\gamma$-quanta prevail, the cluster Compton suppressed $\gamma$-spectrometer proves to be a more effective tool.

The measured characteristics of the phoswich detector cluster presented in this work, such as the suppression factor of the Compton component of the energy spectrum <AC(CsI)>$\simeq$ 53.3\%, energy resolution ($\Delta E$(FWHM)$\simeq$ 30 keV), are sufficient for effectively solving a number of precision $\gamma$-spectroscopy tasks in the region of light nuclei of neutron-rich isotopes Be, B, C, N.

\subsection*{\textbf{Neutron registration efficiency by the phoswich detector cluster}}

The efficiency of neutron registration by the CeBr$_{3}$ part of the phoswich detector in the energy range $E_n$ = 1.4–5.6 MeV was measured using the tagged particle method. A $^{239}$Pu/$^{9}$Be source was used as the neutron source, which is characterized by the fact that in ~60\% of neutron emission acts, the process proceeds via the $^{9}$Be($\alpha$,n)$^{12}$C$^{*}$ channel and is accompanied by the emission of $\gamma$-quanta $E\gamma$ = 4.44 MeV.

Neutrons were identified by coincidences with $\gamma$-quanta $E\gamma$ = 4.44 MeV, which are emitted during the decay of the first excited state of the $^{12}$C$^{}$ nucleus populated in $^{13}$C$^{}\rightarrow n+^{12}$C$^{*}$.
Time-of-flight method(TOF) was used for identification and neutron energy determination.

Recording of an event by the VME data acquisition system was triggered by a logical signal from the trigger CeBr${3}$ detector registering $\gamma$-quanta $E\gamma$ = 4.44 MeV.
Registration of a $\gamma$-quanta $E\gamma$ = 4.44 MeV in the full absorption peak in the trigger detector corresponded to an act of isotropic neutron emission with a known energy distribution \cite{scherzinger2017comparison}.
Events in which the trigger detector registered $\gamma$-quanta $E\gamma$ = 4.44 MeV in the single and double escape peaks of annihilation quanta corresponded to the emission of both neutrons from the source and $\gamma$-quanta $E\gamma$ = 511 keV from the trigger detector.
Based on the known distance between the trigger detector and the phoswich detectors and the position of the $\gamma$-$\gamma$ coincidence peak, the zero point of the time scale was determined, used to calculate the kinetic energy of neutrons $E_n$ from the TOF time.

\begin{figure}[H]
\centering
\begin{minipage}{0.52\textwidth}
	\centering
	\includegraphics*[width=\textwidth]{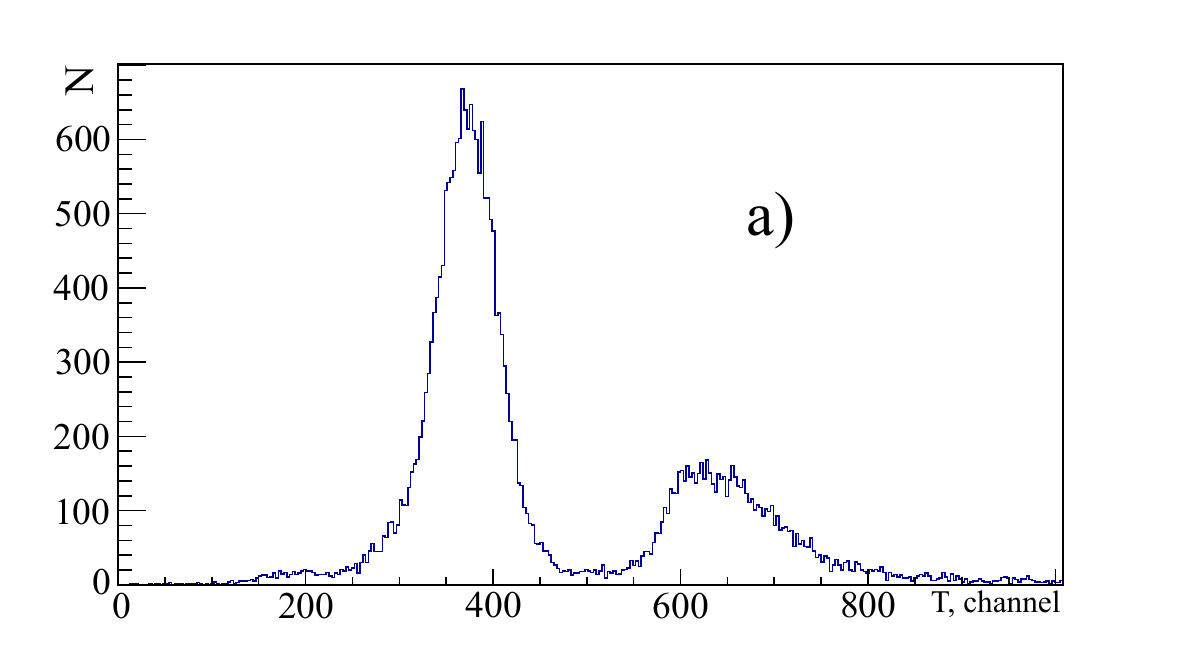}
\end{minipage}
\hfill
\begin{minipage}{0.45\textwidth}
	\centering
	\includegraphics*[width=\textwidth]{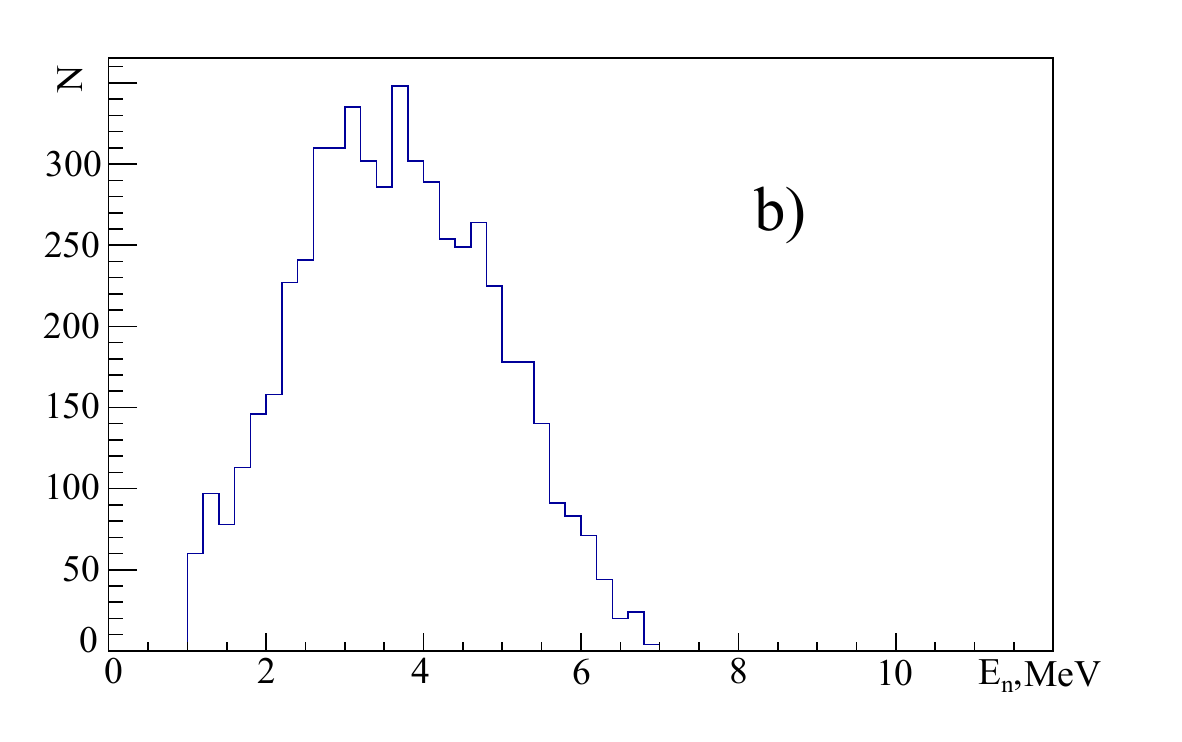}
\end{minipage}
\caption{\it\centering{ a) Time spectrum of $\gamma$-$\gamma$ and $\gamma-n$ coincidence of particles emitted from a $^{239}$Pu/$^{9}$Be source; b) Energy spectrum of neutrons emitted from the $^{239}$Pu/$^{9}$Be source in coincidence with $\gamma$-quanta $E\gamma$ = 4.44 MeV.}}
\label{fig:5}
\end{figure}

\begin{figure} [H]
	\centering
	\includegraphics[width=0.7\linewidth]{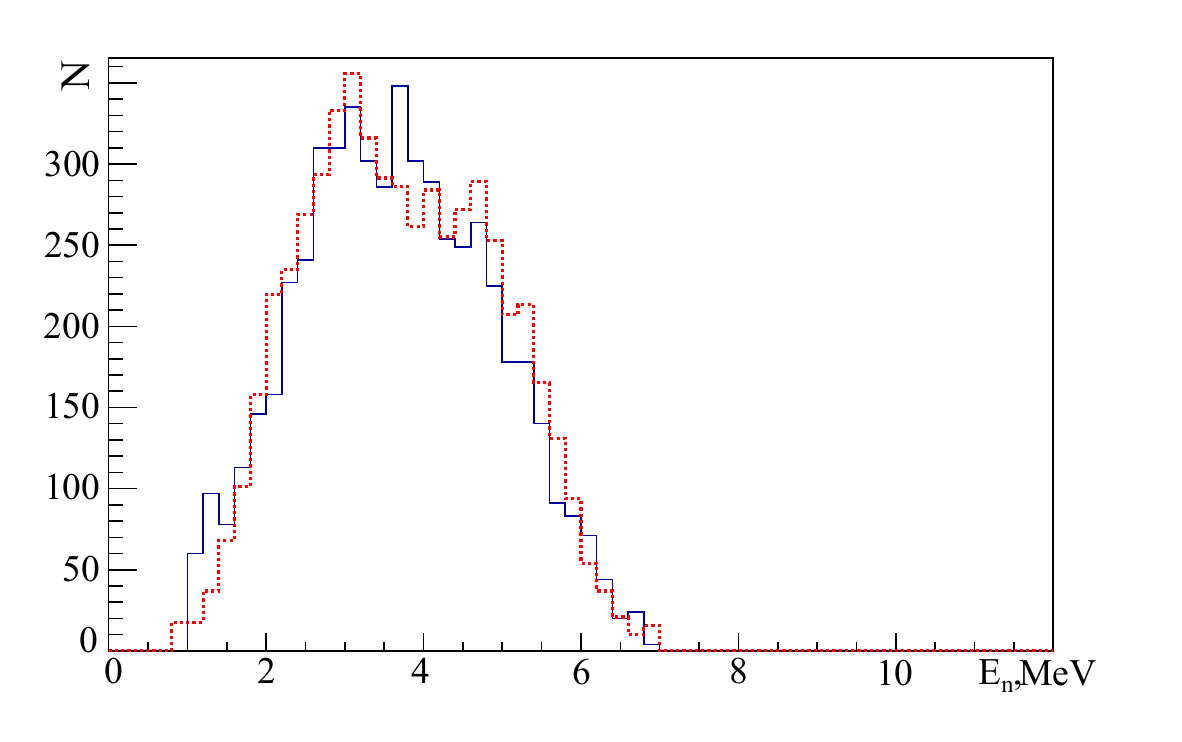}
	\caption{\it\centering{Energy spectrum of neutrons from this work (solid histogram). Neutron spectrum from \cite{scherzinger2017comparison} (dashed histogram).}}
	\label{fig:6}
\end{figure}

The time spectrum of $\gamma-\gamma$ and $\gamma-n$ coincidences of particles emitted from the $^{239}$Pu/$^{9}$Be source is shown in Fig. \ref{fig:5} (a).
The energy spectrum of neutrons registered in coincidence with $\gamma$-quanta from the $^{239}$Pu/$^{9}$Be source is shown in Fig. \ref{fig:5} (b). Neutron energy $E_n$ was calculated using the non-relativistic formula $E_n = \frac{1}{2}m_nv^2$ based on time-of-flight data. The registration threshold was 60 keV on the $\gamma$-scale.
The energy dependence of the neutron registration efficiency in the range E$_{n}$ = 1.4$\div$5.6 MeV was determined by comparing the measured neutron energy spectrum and the spectrum from the work \cite{scherzinger2017comparison}, normalized to the same number of events, see Fig. \ref{fig:6}.
Fig.\ref{fig:6} shows the energy spectra of neutrons from the $^{239}$Pu/$^{9}$Be source, obtained in coincidence with $\gamma$-quanta in this work (solid histogram) and normalized to the same number of events from the work \cite{scherzinger2017comparison} (dashed histogram).

\begin{figure}[H]
	\centering
	\begin{minipage}{0.49\textwidth}
		\centering
		\includegraphics*[width=\textwidth]{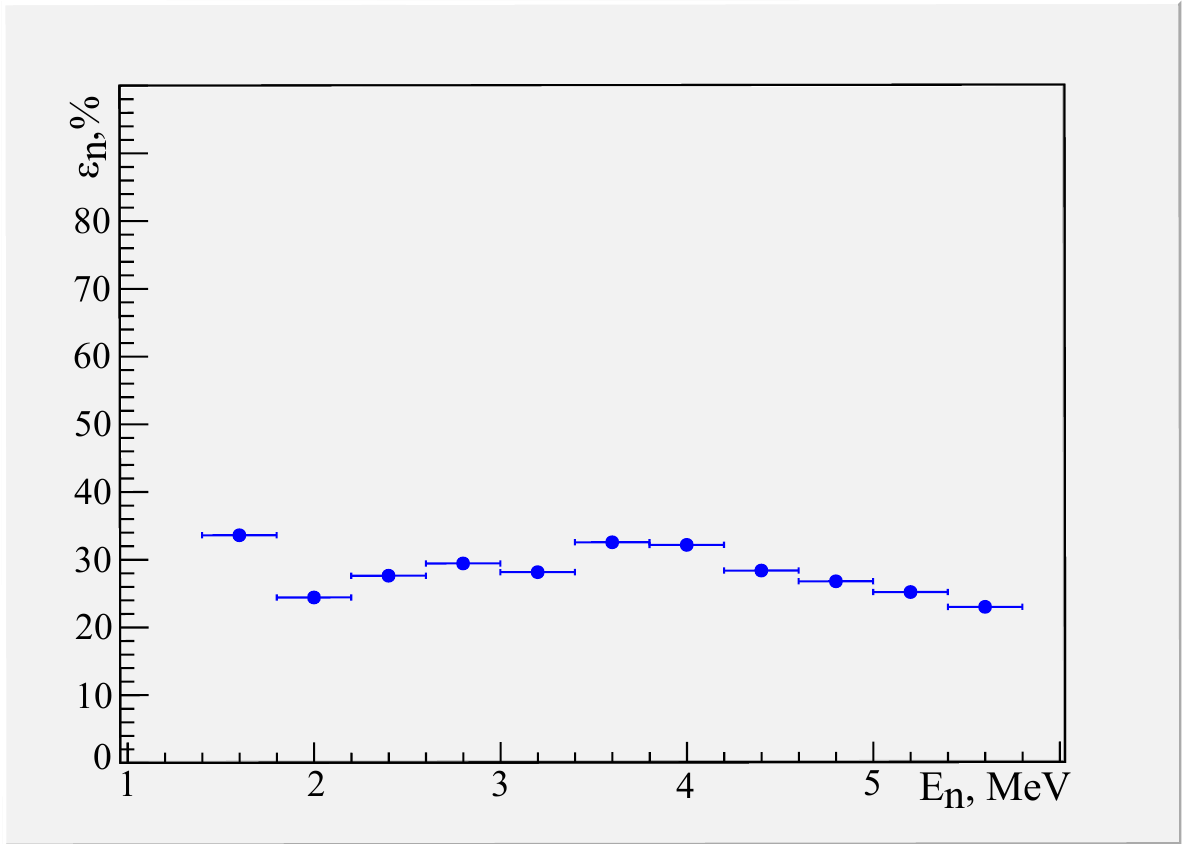}
	\end{minipage}
	\hfill
	\begin{minipage}{0.49\textwidth}
		\centering
		\includegraphics*[width=\textwidth]{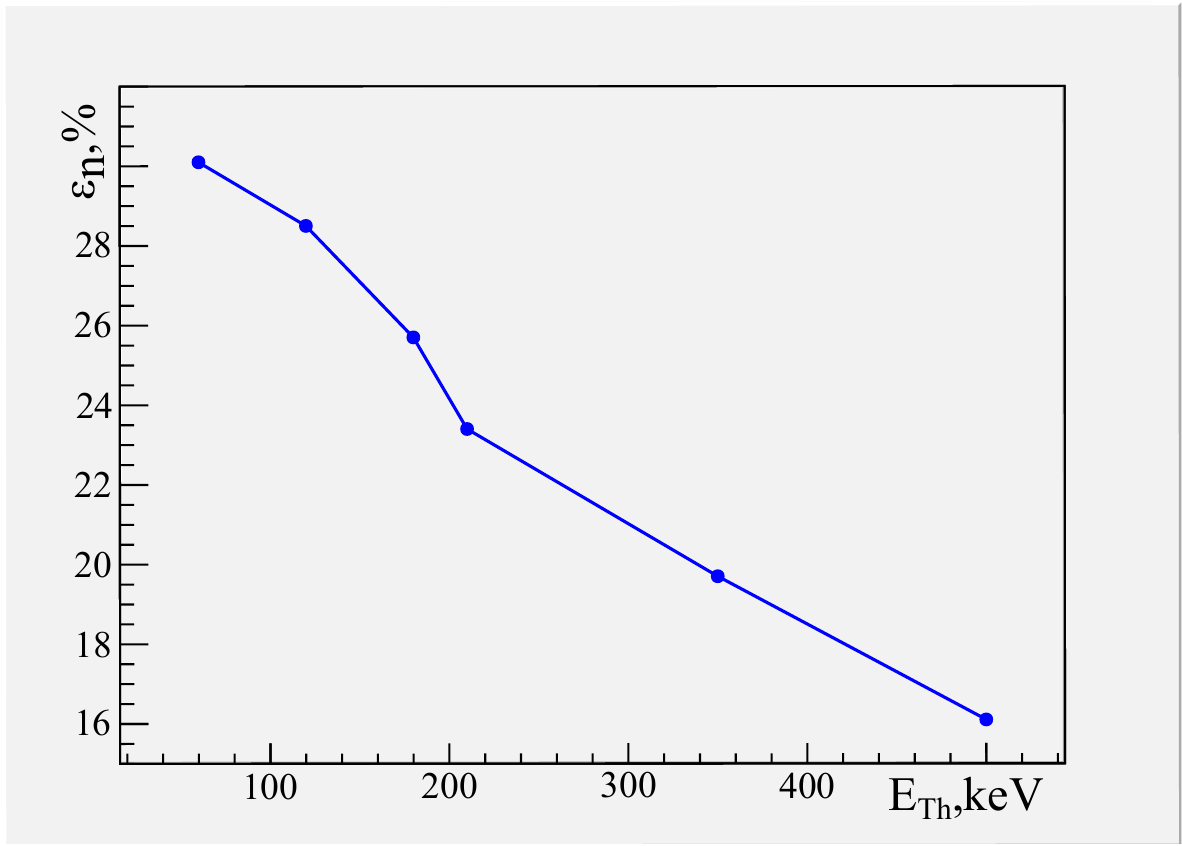}
	\end{minipage}
	\caption{\it\centering{ (Left) Energy dependence of the neutron registration efficiency $\epsilon_n$($E_n$)\% by the CeBr$3$ scintillator of the phoswich detector at a registration threshold $E{th}$ = 60 keV. (Right) Dependence of the average registration efficiency <$\epsilon_n$>\% on the registration threshold value $E_{th}$.}}
	\label{fig:7}
\end{figure}

The energy dependence of the neutron registration efficiency by the CeBr$3$ part of the phoswich detector cluster at different registration threshold values was obtained as a result of analyzing the neutron energy spectra. Fig.\ref{fig:7} Left: shows the energy dependence of the neutron registration efficiency $\epsilon_n$\% at a registration threshold $E{th}$ = 60 keV. Right: Dependence of the average registration efficiency $\bar{\epsilon_n}$\% in the energy range $E_n$= 1.4$\div$5.6 MeV on the registration threshold value E$_{th}$ (the abscissa axis shows the registration threshold on the energy $\gamma$ -scale).

The highest value $\bar{\epsilon_n}$ $\simeq$ 30\% was obtained at an energy registration threshold E$_{th}$= 60 keV.
These results showed that the CeBr$_{3}$ scintillation detector allows for neutron registration with relatively high efficiency, and also confirms the main advantage of scintillators of this type — the low dependence of efficiency on the energy of the registered neutron over a wide energy range $E_n$ = 1.4 $\div$ 5.6 MeV.

\section*{\textbf{Conclusion}}
As a result of the measurements, the main characteristics of the spectrometer based on the 9$\times$CeBr$_{3}$-NaI(Tl) cluster with an Compton suppressed shield block of 4$\times$CsI(Tl) were determined.

The measured characteristics of the spectrometer for $\gamma$-radiation registration (energy resolution, registration efficiency and full absorption peak efficiency, as well as suppression factors of the Compton part of the $\gamma$-quantum energy spectrum) demonstrated its high effectiveness.

Equipping the cluster with four CsI(Tl) detectors allowed increasing the suppression factor of the Compton part of the spectrum by a maximum of $\sim$1.5 times.

The maximum suppression factor value was $\sim$65$\%$ for gamma radiation from the $^{60}$Co source.

 The dependence of registration efficiency on energy $\varepsilon$(E${n}$) in the range E${n}$=1.4$\div$5.6 MeV when registering neutrons from the $^{239}$Pu/$^{9}$Be source was obtained. The efficiency demonstrates a flat dependence on energy with an average value $\varepsilon$(E$_{n}$)=$\simeq$30$\%$ for the entire range, which agrees with data previously obtained for inorganic scintillators \cite{matulewicz1989response}.

Additionally, the dependence of efficiency on the energy registration threshold was investigated, demonstrating the necessity of operating at minimal thresholds.
The authors thank the Russian Science Foundation for financial support (Project N 24-22-00117).
	\printbibliography

@article{moses2005potential,
	title={Potential for RbGd2Br7: Ce, LaBr3: Ce, LaBr3: Ce, and LuI3: Ce in nuclear medical imaging},
	author={Moses, William W and Shah, Kanai S},
	journal={Nuclear Instruments and Methods in Physics Research Section A: Accelerators, Spectrometers, Detectors and Associated Equipment},
	volume={537},
	number={1-2},
	pages={317--320},
	year={2005},
	publisher={Elsevier}
}

@article{yeom2013first,
	title={First performance results of Ce: GAGG scintillation crystals with silicon photomultipliers},
	author={Yeom, Jung Yeol and Yamamoto, Seiichi and Derenzo, Stephen E and Spanoudaki, Virginia Ch and Kamada, Kei and Endo, Takanori and Levin, Craig S},
	journal={IEEE transactions on nuclear science},
	volume={60},
	number={2},
	pages={988--992},
	year={2013},
	publisher={IEEE}
}

@article{qin2020characteristics,
	title={Characteristics and time resolutions of two CeBr3 gamma-ray spectrometers},
	author={Qin, Jianguo and Lai, Caifeng and Xiao, Jun and Lu, Xinxin and Zhu, Tonghua and Liu, Rong and Ye, Bangjiao},
	journal={Radiation Detection Technology and Methods},
	volume={4},
	number={3},
	pages={327--336},
	year={2020},
	publisher={Springer}
}

@article{challan2013gamma,
  title={Gamma-ray efficiency of a HPGe detector as a function of energy and geometry},
  author={Challan, Mohsen B},
  journal={Applied Radiation and Isotopes},
  volume={82},
  pages={166--169},
  year={2013},
  publisher={Elsevier}
}

@article{sivavcek2020multi,
	title={MULTI-2, a 4$\pi$ spectrometer for total reaction cross section measurements},
	author={Siv{\'a}{\v{c}}ek, I and Penionzhkevich, Yu E and Sobolev, Yu G and Stukalov, SS},
	journal={Nuclear Instruments and Methods in Physics Research Section A: Accelerators, Spectrometers, Detectors and Associated Equipment},
	volume={976},
	pages={164255},
	year={2020},
	publisher={Elsevier}
}

@article{wasilewska2022gamma,
	title={$\gamma$ decay to the ground state from the excitations above the neutron threshold in the Pb 208 (p, p' $\gamma$) reaction at 85 MeV},
	author={Wasilewska, B and Kmiecik, M and Ciema{\l}a, M and Maj, A and Crespi, FCL and Bracco, A and Harakeh, MN and Bednarczyk, P and Bottoni, S and Brambilla, S and others},
	journal={Physical Review C},
	volume={105},
	number={1},
	pages={014310},
	year={2022},
	publisher={APS}
}

@article{li2025beta,
	title={$\beta$-delayed spectroscopy of Ge 48 80 and competition between Gamow-Teller and first-forbidden transitions in Ga 49 80 g+ m $\beta$ decay},
	author={Li, R and Verney, D and Matea, I and Harakeh, MN and Delafosse, C and Didierjean, F and Ebata, S and Ayoubi, LA and Al Falou, H and Benzoni, G and others},
	journal={Physical Review C},
	volume={112},
	number={4},
	pages={044306},
	year={2025},
	publisher={APS}
}

@article{matulewicz1989response,
	title={Response of BaF$_{2}$, CsI (Tl) and Pb-glass detectors to neutrons below 22 MeV},
	author={Matulewicz, T and Grosse, E and Emling, H and Grein, H and Kulessa, R and Baumann, FM and Domogala, G and Freiesleben, H},
	journal={Nuclear Instruments and Methods in Physics Research Section A: Accelerators, Spectrometers, Detectors and Associated Equipment},
	volume={274},
	number={3},
	pages={501--506},
	year={1989},
	publisher={Elsevier}
}

@article{quarati2013scintillation,
	title={Scintillation and detection characteristics of high-sensitivity CeBr$_{3}$ $\gamma$-ray spectrometers},
	author={Quarati, FGA and Dorenbos, P and Van Der Biezen, J and Owens, Alan and Selle, M and Parthier, L and Schotanus, P},
	journal={Nuclear Instruments and Methods in Physics Research Section A: Accelerators, Spectrometers, Detectors and Associated Equipment},
	volume={729},
	pages={596--604},
	year={2013},
	publisher={Elsevier}
}

@article{britton2014monte,
	title={Monte-Carlo optimisation of a Compton suppression system for use with a broad-energy HPGe detector},
	author={Britton, R and Burnett, JL and Davies, AV and Regan, PH},
	journal={Nuclear Instruments and Methods in Physics Research Section A: Accelerators, Spectrometers, Detectors and Associated Equipment},
	volume={762},
	pages={42--53},
	year={2014},
	publisher={Elsevier}
}

@article{scherzinger2017comparison,
	title={A comparison of untagged gamma-ray and tagged-neutron yields from $^{241}$AmBe and $^{238}$PuBe sources},
	author={Scherzinger, Julius and Al Jebali, Ramsey and Annand, JRM and Fissum, KG and Hall-Wilton, Richard and Koufigar, Sharareh and Mauritzson, Nicholai and Messi, Francesco and Perrey, Hanno and Rofors, Emil},
	journal={Applied Radiation and Isotopes},
	volume={127},
	pages={98--102},
	year={2017},
	publisher={Elsevier}
}

@manual{ssyl4,
	title={https://www.mesytec.com/}
}
\end{document}